\documentstyle[twoside]{article}

\catcode`\@=11
\long\def\@makefntext#1{
\protect\noindent \hbox to 3.2pt {\hskip-.9pt
$^{{\eightrm\@thefnmark}}$\hfil}#1\hfill}               

\def\thefootnote{\fnsymbol{footnote}}
\def\@makefnmark{\hbox to 0pt{$^{\@thefnmark}$\hss}}    

\def\ps@myheadings{\let\@mkboth\@gobbletwo
\def\@oddhead{\hbox{}
\rightmark\hfil\eightrm\thepage}
\def\@oddfoot{}\def\@evenhead{\eightrm\thepage\hfil
\leftmark\hbox{}}\def\@evenfoot{}
\def\sectionmark##1{}\def\subsectionmark##1{}}

\oddsidemargin=\evensidemargin
\renewcommand{\thefootnote}{\fnsymbol{footnote}}

\newcounter{sectionc}\newcounter{subsectionc}\newcounter{subsubsectionc}
\renewcommand{\section}[1] {\vspace{12pt}\addtocounter{sectionc}{1}
\setcounter{subsectionc}{0}\setcounter{subsubsectionc}{0}\noindent
        {\tenbf\thesectionc. #1}\par\vspace{5pt}}
\renewcommand{\subsection}[1] {\vspace{12pt}\addtocounter{subsectionc}{1}
        \setcounter{subsubsectionc}{0}\noindent
        {\bf\thesectionc.\thesubsectionc. {\kern1pt \bfit #1}}\par\vspace{5pt}}
\renewcommand{\subsubsection}[1] {\vspace{12pt}\addtocounter{subsubsectionc}{1}
        \noindent{\tenrm\thesectionc.\thesubsectionc.\thesubsubsectionc.
        {\kern1pt \tenit #1}}\par\vspace{5pt}}
\newcommand{\nonumsection}[1] {\vspace{12pt}\noindent{\tenbf #1}
        \par\vspace{5pt}}

\newcounter{appendixc}
\newcounter{subappendixc}[appendixc]
\newcounter{subsubappendixc}[subappendixc]
\renewcommand{\thesubappendixc}{\Alph{appendixc}.\arabic{subappendixc}}
\renewcommand{\thesubsubappendixc}
        {\Alph{appendixc}.\arabic{subappendixc}.\arabic{subsubappendixc}}

\renewcommand{\appendix}[1] {\vspace{12pt}
        \refstepcounter{appendixc}
        \setcounter{figure}{0}
        \setcounter{table}{0}
        \setcounter{lemma}{0}
        \setcounter{theorem}{0}
        \setcounter{corollary}{0}
        \setcounter{definition}{0}
        \setcounter{equation}{0}
        \renewcommand{\thefigure}{\Alph{appendixc}.\arabic{figure}}
        \renewcommand{\thetable}{\Alph{appendixc}.\arabic{table}}
        \renewcommand{\theappendixc}{\Alph{appendixc}}
        \renewcommand{\thelemma}{\Alph{appendixc}.\arabic{lemma}}
        \renewcommand{\thetheorem}{\Alph{appendixc}.\arabic{theorem}}
        \renewcommand{\thedefinition}{\Alph{appendixc}.\arabic{definition}}
        \renewcommand{\thecorollary}{\Alph{appendixc}.\arabic{corollary}}
        \renewcommand{\theequation}{\Alph{appendixc}.\arabic{equation}}
        \noindent{\tenbf Appendix \theappendixc #1}\par\vspace{5pt}}
\newcommand{\subappendix}[1] {\vspace{12pt}
        \refstepcounter{subappendixc}
        \noindent{\bf Appendix \thesubappendixc. {\kern1pt \bfit #1}}
        \par\vspace{5pt}}
\newcommand{\subsubappendix}[1] {\vspace{12pt}
        \refstepcounter{subsubappendixc}
        \noindent{\rm Appendix \thesubsubappendixc. {\kern1pt \tenit #1}}
        \par\vspace{5pt}}

\newcommand{\textlineskip}{\baselineskip=13pt}
\newcommand{\smalllineskip}{\baselineskip=10pt}

\def\eightcirc{
\begin{picture}(0,0)
\put(4.4,1.8){\circle{6.5}}
\end{picture}}
\def\eightcopyright{\eightcirc\kern2.7pt\hbox{\eightrm c}}

\newcommand{\copyrightheading}[1]
        {\vspace*{-2.5cm}\smalllineskip{\flushleft
        {\footnotesize $\eightcopyright$\, World Scientific Publishing
         Company}\\
         }}

\newcommand{\publisher}[2]{{\begin{center}\footnotesize\smalllineskip
        Received #1\\
        Revised #2
        \end{center}
        }}

\def\abstracts#1#2#3{{
        \centering{\begin{minipage}{4.5in}\baselineskip=10pt\footnotesize
        \parindent=0pt #1\par
        \parindent=15pt #2\par
        \parindent=15pt #3
        \end{minipage}}\par}}

\newcommand{\bibit}{\nineit}
\newcommand{\bibbf}{\ninebf}
\renewenvironment{thebibliography}[1]
        {\frenchspacing
         \ninerm\baselineskip=11pt
         \begin{list}{\arabic{enumi}.}
        {\usecounter{enumi}\setlength{\parsep}{0pt}
         \setlength{\leftmargin 12.7pt}{\rightmargin 0pt} 
         \setlength{\itemsep}{0pt} \settowidth
        {\labelwidth}{#1.}\sloppy}}{\end{list}}

\newcounter{itemlistc}
\newcounter{romanlistc}
\newcounter{alphlistc}
\newcounter{arabiclistc}

\newcommand{\fcaption}[1]{
        \refstepcounter{figure}
        \setbox\@tempboxa = \hbox{\footnotesize Fig.~\thefigure. #1}
        \ifdim \wd\@tempboxa > 5in
           {\begin{center}
        \parbox{5in}{\footnotesize\smalllineskip Fig.~\thefigure. #1}
            \end{center}}
        \else
             {\begin{center}
             {\footnotesize Fig.~\thefigure. #1}
              \end{center}}
        \fi}

\newcommand{\tcaption}[1]{
        \refstepcounter{table}
        \setbox\@tempboxa = \hbox{\footnotesize Table~\thetable. #1}
        \ifdim \wd\@tempboxa > 5in
           {\begin{center}
        \parbox{5in}{\footnotesize\smalllineskip Table~\thetable. #1}
            \end{center}}
        \else
             {\begin{center}
             {\footnotesize Table~\thetable. #1}
              \end{center}}
        \fi}

\def\@citex[#1]#2{\if@filesw\immediate\write\@auxout
        {\string\citation{#2}}\fi
\def\@citea{}\@cite{\@for\@citeb:=#2\do
        {\@citea\def\@citea{,}\@ifundefined
        {b@\@citeb}{{\bf ?}\@warning
        {Citation `\@citeb' on page \thepage \space undefined}}
        {\csname b@\@citeb\endcsname}}}{#1}}

\newif\if@cghi
\def\cite{\@cghitrue\@ifnextchar [{\@tempswatrue
        \@citex}{\@tempswafalse\@citex[]}}
\def\citelow{\@cghifalse\@ifnextchar [{\@tempswatrue
        \@citex}{\@tempswafalse\@citex[]}}
\def\@cite#1#2{{$\null^{#1}$\if@tempswa\typeout
        {IJCGA warning: optional citation argument
        ignored: `#2'} \fi}}

\def\pmb#1{\setbox0=\hbox{#1}
        \kern-.025em\copy0\kern-\wd0
        \kern.05em\copy0\kern-\wd0
        \kern-.025em\raise.0433em\box0}

\def\fnt#1#2{\footnotetext{\kern-.3em
        {$^{\mbox{\scriptsize #1}}$}{#2}}}

\def\fpage#1{\begingroup
\voffset=.3in
\thispagestyle{empty}\begin{table}[b]\centerline{\footnotesize #1}
        \end{table}\endgroup}

\def\runninghead#1#2{\pagestyle{myheadings}
\markboth{{\protect\footnotesize\it{\quad #1}}\hfill}
{\hfill{\protect\footnotesize\it{#2\quad}}}}
\font\tenrm=cmr10
\font\tenit=cmti10
\font\tenbf=cmbx10
\font\bfit=cmti10 at 10pt
\font\ninerm=cmr9
\font\nineit=cmti9
\font\ninebf=cmbx9
\font\eightrm=cmr8

\def\qed{\hbox{${\vcenter{\vbox{                        
   \hrule height 0.4pt\hbox{\vrule width 0.4pt height 6pt
   \kern5pt\vrule width 0.4pt}\hrule height 0.4pt}}}$}}

\renewcommand{\thefootnote}{\fnsymbol{footnote}}        

\def\bsc{{\sc a\kern-6.4pt\sc a\kern-6.4pt\sc a}}       
\def\bflatex{\bf L\kern-.30em\raise.3ex\hbox{\bsc}\kern-.14em
T\kern-.1667em\lower.7ex\hbox{E}\kern-.125em X}

\def\selecR{\tilde e_{\rm R}}
\def\selecL{\tilde e_{\rm L}}

\def\lsp{{\tilde\chi^0}_1}
\def\cinolm{\tilde{\chi_{\rm 1}}^-}
\def\cinolp{\tilde{\chi_{\rm 1}}^+}
\def\photino{\tilde\gamma}
\begin{document}

\runninghead{A System for the Automatic Computation}
{A System for the Automatic Computation}

\normalsize\textlineskip
\thispagestyle{empty}
\setcounter{page}{1}

\copyrightheading{}                     

\vspace*{0.88truein}

\fpage{1}
\centerline{\bf A SYSTEM FOR THE AUTOMATIC COMPUTATION OF}
\vspace*{0.035truein}
\centerline{\bf CROSS-SECTIONS INCLUDING SUSY PARTICLES}
\vspace*{0.37truein}
\centerline{\footnotesize MASATO JIMBO}
\vspace*{0.015truein}
\centerline{\footnotesize\it  Computer Science Laboratory, Tokyo Management
College, 625-1 Futamata}
\baselineskip=10pt
\centerline{\footnotesize\it Ichikawa, Chiba 272, Japan}
\vspace*{10pt}
\centerline{\footnotesize TADASHI KON}
\vspace*{0.015truein}
\centerline{\footnotesize\it Faculty of Engineering, Seikei University,
3-3-1 Kichijoji-kita}
\baselineskip=10pt
\centerline{\footnotesize\it Musashino, Tokyo 180, Japan}
\vspace*{10pt}
\centerline{\footnotesize HIDEKAZU TANAKA}
\vspace*{0.015truein}
\centerline{\footnotesize\it Faculty of Science, Rikkyo University,
3-34-1 Nishi-ikebukuro}
\baselineskip=10pt
\centerline{\footnotesize\it Toshima, Tokyo 171, Japan}
\vspace*{10pt}
\centerline{\footnotesize TOSHIAKI KANEKO\footnote{Present address:
L.A.P.P., F-74019 Annecy le vieux Cedex, France}}
\vspace*{0.015truein}
\centerline{\footnotesize\it Faculty of General Education, Meiji-gakuin
University, 1518 Kamikurata}
\baselineskip=10pt
\centerline{\footnotesize\it Totsuka, Yokohama 241, Japan}
\vspace*{10pt}
\centerline{\footnotesize MINAMI-TATEYA COLLABORATION}
\vspace*{0.225truein}
\publisher{(received date)}{(revised date)}

\vspace*{0.21truein}
\abstracts{ We introduce a new method to treat Majorana fermions on the GRACE
system, which has already been developed for the computation of the matrix
elements for the processes of the standard model.  In the standard model, we
already have included such particles as Dirac fermions, gauge bosons and
scalar bosons in the system.   On the other hand, there are four Majorana
fermions called neutralinos in the minimal SUSY standard model (MSSM).
In consequence, we have constructed a system for the automatic computation of
cross-sections for the processes of the MSSM.   It is remarkable that our
system is also applicable for another model including Majorana fermions once
the definition of the model file is given.
}{}{}



\vspace*{1pt}\textlineskip      
\section{Introduction}          
\vspace*{-0.5pt}
\noindent
The quest of new particles is one of the most important aim of the present
high-energy physics.  Though there are many theories which predict undiscovered
particles, supersymmetric (SUSY) theories are fascinating ones because of
the beautiful symmetry between bosons and fermions at the unification-energy
scale.  It, however, is a broken symmetry at the electroweak-energy scale.
The relic of SUSY is expected to remain as a rich spectrum of SUSY particles,
partners of usual matter fermions, gauge bosons and Higgs scalars, named
sfermions, gauginos and higgsinos, respectively.$^1$

Since there exist so many particles and their interactions, it is a skilled job
to calculate the cross-sections for the processes with the final 3-body or
more.  We have already known within the standard model that the calculation of
the helicity amplitudes is more advantageous to such a case than that of the
traces for the gamma matrices with REDUCE.$^{2,3}$  The program package
CHANEL$~^4$ is one of the utilities for the numerical calculation of the
helicity amplitudes.

It, however, is also hard work to construct a program with many subroutine
calls of CHANEL by hand.  Thus we need a more convenient way to carry out such
a work.  Several groups have started independently to develop computer systems
which automate the perturbative calculation in the standard model with
different methods.$^{5-8}$  The GRACE system,$^5$ which automatically
generates the source code for CHANEL, is one of the solutions.  The system also
includes the interface and the library of CHANEL, and the multi-dimensional
integration and event-generation package BASES/SPRING v5.1.$^9$

In the SUSY models, there exist Majorana fermions as the neutral gauginos
and higgsinos, which become the mixed states called neutralinos.  Since
anti-particles of Majorana fermions are themselves, there exists so-called
`Majorana-flip', the transition between particle and anti-particle.  This has
been the most important problem which we should solve when we realize the
automatic system for computation of the SUSY processes.

In a recent work,$^{10,11}$ we developed an algorithm to treat Majorana
fermions in CHANEL.  In the standard model, we already have such particles as
Dirac fermions, gauge bosons and scalar bosons in the GRACE system.  Thus we
have constructed an automatic system for the computation of the SUSY processes
by the algorithm above in the GRACE system.

\textheight=7.8truein
\setcounter{footnote}{0}
\renewcommand{\thefootnote}{\alph{footnote}}

\section{Majorana fermions into new GRACE}
\noindent
The GRACE system in the public ftp-site is for the automatic computation of the
cross-sections for the processes within the standard model.$^{12}$  The GRACE
system has become more flexible for the extension in the new version called
`{\bf grc}',$^{13}$ which includes a new graph-generation package.$^{14}$  With
this package, every graph can be generated based on a user-defined model.  It
is necessary for us to make the library and the interface of CHANEL and the
model file for including the SUSY particles.

The method of computation in the program package CHANEL is as follows:
\begin{enumerate}
 \item To divide a helicity amplitude into vertex amplitudes.
 \item To calculate each vertex amplitude numerically as a complex number.
 \item To reconstruct of them with the polarization sum, and calculate
 the helicity amplitudes numerically.
\end{enumerate}
The merit of this method is that the extension of the package is easy,
and that each vertex can be defined only by the type of concerned particles.

Here we adopt an algorithm in Ref. 10 and 11 for the implementation of the
embedding Majorana fermions in CHANEL as follows:
\begin{itemize}
 \item \underline{\bf policy}
 \begin{enumerate}
  \item To calculate a helicity amplitude numerically.
  \item To replace each propagator by wave functions or polarization vectors,
  and calculate vertex amplitudes.
  \item \underline{\bf Not to} move charge-conjugation matrices.
 \end{enumerate}
 \item \underline{\bf method}
 \begin{enumerate}
  \item To choose a direction on a fermion line.
  \item To put wave functions, vertices and propagators along the direction
  in such a way:
  \begin{itemize}
   \item[~i)] To take the transpose for the reverse direction of fermions
   \item[ii)] To use the propagator with the charge-conjugation matrix\\
   for the Majorana-flipped one.
  \end{itemize}
 \end{enumerate}
\end{itemize}
As a result, the kinds of the Dirac-Majorana-scalar vertices are limited to
four types:
\begin{eqnarray}
 J_{1~{h_1}{h_2}}^{[{\rm S}]{\rho_1}{\rho_2}} & = &
  \overline{U}^{\rho_1}({h_1},{p_1},{m_1}) \Gamma
  U^{\rho_2}({h_2},{p_2},{m_2})~~, \\
 J_{2~{h_1}{h_2}}^{[{\rm S}]{\rho_1}{\rho_2}} & = &
  U^{{\rho_1}\rm T}({h_1},{p_1},{m_1}) \Gamma
  \overline{U}^{{\rho_2}\rm T}({h_2},{p_2},{m_2})~~, \\
 J_{3~{h_1}{h_2}}^{[{\rm S}]{\rho_1}{\rho_2}} & = &
  \overline{U}^{\rho_1}({h_1},{p_1},{m_1}) C^{\rm T} \Gamma^{\rm T}
  \overline{U}^{{\rho_2}\rm T}({h_2},{p_2},{m_2})~~, \\
 J_{4~{h_1}{h_2}}^{[{\rm S}]{\rho_1}{\rho_2}} & = &
  U^{{\rho_1}\rm T}({h_1},{p_1},{m_1}) \Gamma^{\rm T} C^{-1}
  U^{\rho_2}({h_2},{p_2},{m_2})~~,
\end{eqnarray}
where $U$'s denote wave functions, and $C$ is the charge-conjugation matrix.
The symbol $\Gamma$ stands for the scalar vertex such as
 \[ \Gamma = A_{\rm L}\cdot{{1 - \gamma}\over{2}} +
      A_{\rm R}\cdot{{1 + \gamma}\over{2}} ~~. \]
The vertices $J_{2}^{[{\rm S}]} \sim J_{4}^{[{\rm S}]}$ are related to the
vertex $J_{1}^{[{\rm S}]}$ which has been already defined as the
Dirac-Dirac-scalar vertex in the subroutine FFS of CHANEL.  The relations among
the vertices are as follows:
\begin{eqnarray}
 J_{1~{h_1}{h_2}}^{[{\rm S}]{\rho_1}{\rho_2}} \qquad & & \rightarrow
 {\rm FFS}~~,\\
 J_{2~{h_1}{h_2}}^{[{\rm S}]{\rho_1}{\rho_2}} \quad = &
  -J_{1~{h_1}{h_2}}^{[{\rm S}]{-\rho_1}{-\rho_2}}~ & \rightarrow
  {\rm FFST}~~,\\
 J_{3~{h_1}{h_2}}^{[{\rm S}]{\rho_1}{\rho_2}} \quad = &
  -J_{1~{h_1}{h_2}}^{[{\rm S}]{\rho_1}{-\rho_2}} \quad & \rightarrow
  {\rm FFCS}~~,\\
 J_{4~{h_1}{h_2}}^{[{\rm S}]{\rho_1}{\rho_2}} \quad = &
  -J_{1~{h_1}{h_2}}^{[{\rm S}]{-\rho_1}{\rho_2}} \quad & \rightarrow
  {\rm FFSC}~~,
\end{eqnarray}
Thus we have built three new subroutines FFST, FFCS and FFSC.  We have
performed the installation of the subroutines above with the interface on the
new GRACE system.

\section{Tests for the system}
\noindent
At the start for the check of our system, we have written the model file
of SUSY QED.  In this case, there is only one Majorana fermion, photino.
Next we have extended the model file and the definition file of couplings for
the MSSM.  The tests have been performed by the exact calculations with the two
methods, our system and REDUCE, in such a manner:
\begin{enumerate}
  \item To calculate the differential cross-sections at a point of the
  phase space in the two methods with GRACE and REDUCE.
  \item To calculate the differential cross-sections over the
  phase space with REDUCE.
  \item To integrate the differential cross-sections over the
  phase space in the two methods with GRACE and REDUCE.
\end{enumerate}
With the GRACE system, we can get the differential cross-sections and the
scattered plots by one time of the integration step with BASES.  For writing
REDUCE sources, we use the different method to treat Majorana fermions in
Ref. 15.  In Table I, the tested processes are shown as a list.

\begin{table}[htbp]
\tcaption{The list of the tested processes.}
\centerline{\footnotesize\smalllineskip
\begin{tabular}{llclc}
\hline
 Process &  & Number of diagrams & Comment & Check \\ \hline
{\footnotesize\bf SUSY} & {\footnotesize\bf QED} & & & \\ \hline
$e^- e^- \quad\rightarrow$ & $\selecR^- \selecR^-$ & 2 & Majorana-flip & OK \\
 & $\selecL^- \selecL^-$ & 2 & in internal lines & OK \\
 & $\selecR^- \selecL^-$ & 2 & & OK \\ \hline
$e^- e^+ \quad\rightarrow$ & $\selecR^- \selecR^+$ & 2 & Including pair & OK \\
 & $\selecL^- \selecL^+$ & 2 & annihilation & OK \\ \hline
$e^- e^+ \quad\rightarrow$ & $\selecR^- \selecL^+$ & 1 & Values are & OK \\
 & $\selecR^+ \selecL^-$ & 1 & equal & OK \\  \hline
$e^- e^+ \quad\rightarrow$ & $\photino \photino$ & 4 & F-B symmetric & OK
 \\ \hline
$e^- e^+ \quad\rightarrow$ & $\photino \photino \gamma~$ & 12 & Final 3-body
 & OK \\ \hline
$e^- e^+ \quad\rightarrow$ & $\selecR^- \photino e^+$ & 12 & Everything for
 tests & OK \\ \hline
{\footnotesize\bf MSSM} & & & \\ \hline
$e^- e^- \quad\rightarrow$ & $\selecL^- \selecL^-$ & 8 & 4 Majorana fermions
 & OK \\ \hline
$e^- e^+ \quad\rightarrow$ & $\cinolm \cinolp$ & 3 & & OK \\ \hline
$e^- e^+ \quad\rightarrow$ & $\lsp \lsp \gamma$ & 14 & Final 3-body
 & OK \\ \hline\\
\end{tabular}}
\end{table}

For an example of the tests, we present the results for the process of the
single-selectron production, $e^- e^+ \rightarrow \selecR^- \photino e^+$
within SUSY QED.  In this process, there exist diagrams with Majorana-flips in
internal and external lines, and ones with one-photon exchange in s-channel and
t-channel.$^{16}$  Thus this is the most important process for the test of our
system.  In Fig.~1, we show the angular distribution of the outgoing positron
in this process.  Here we use BASES for the calculation from the REDUCE output.
The result is in beautiful agreement with the value that is obtained by GRACE
at each bin of the histograms.
\begin{figure}[htbp]
\vspace*{13pt}
\centerline{\vbox{\hrule width 7cm height0.001pt}}
\vspace*{1.6truein}             
\centerline{\vbox{\hrule width 7cm height0.001pt}}
\vspace*{13pt}
\hspace*{1.7cm} Fig.~1. Angular distribution of the positron in $e^- e^+
   \rightarrow \selecR^- \photino e^+$.
\end{figure}
Since the two diagrams with the one-photon exchange dominate in this case,
there is a steep peak in the direction of the initial positron.  In such a
case, the equivalent-photon approximation (EPA) works well.$^{17}$  The
comparison between the results of the exact calculation and those of EPA is
shown in Ref. 18.  In the experiment, the single-photon event from $e^- e^+
\rightarrow \lsp \lsp \gamma$ is important for the search of SUSY particles.
We have also calculated this process with our system.$^{19}$

\section{Summary}
\noindent
We introduce a new method to treat Majorana fermions on the GRACE system for
the automatic computation of the matrix elements for the processes of the SUSY
models.  In the first instance, we have constructed the system for the
processes of the SUSY QED because we should test our algorithm for the
simplest case.  Next we have extended the model file and the definition file of
couplings for the MSSM.  The numerical results convince us that our algorithm
is correct.  It is remarkable that our system is also applicable to another
model including Majorana fermions.

\nonumsection{Acknowledgements}
\noindent
This work was supported in part by the Ministry of Education,
Science and Culture, Japan under Grant-in-Aid for International
Scientific Research Program No. 04044158.  Two of the authors (H.T. and M.J.)
have been also indebted to the above-mentioned Ministry under
Grant-in-Aid No.06640411.

\nonumsection{References}

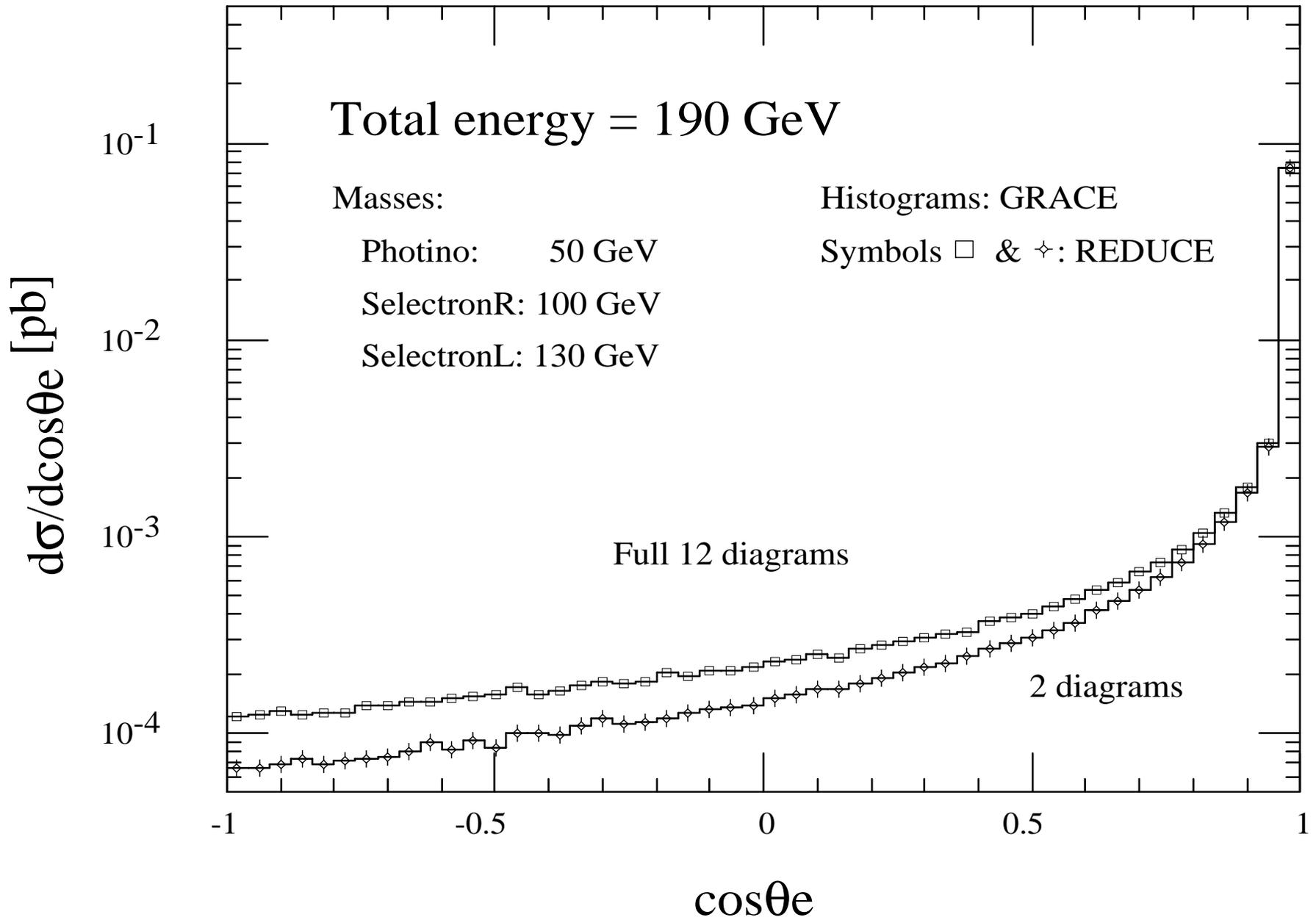
\end{document}